\documentstyle[amsmath,aps,prl,multicol,epsfig,floats,amssymb,
graphics]{revtex}
\def\8{\infty}
\def\oh{\frac{1}{2}}

\def\undertext#1{\vtop{\hbox{#1}\kern 1pt \hrule}}

\def\VEV#1{\left\langle\,#1\,\right\rangle}

\def\dd#1{\frac{d}{d#1}}

\def\be{\begin{equation}}
\def\ee{\end{equation}}
\def\bea{\begin{eqnarray} & &}
\def\eea{\end{eqnarray}}

\def\rf#1{(\ref{#1})}

\def\cH{{\cal H}}
\def\rf#1{(\ref{#1})}
\def\t{\tilde}
\def\cE{{\cal E}}

\begin{document}
\draft
\twocolumn[

\hsize\textwidth\columnwidth\hsize\csname @twocolumnfalse\endcsname

\title{Some generic aspects of bosonic excitations in disordered systems}
\author{V. Gurarie and J. T. Chalker}
\address{Theoretical Physics, Oxford University, 1 Keble Road,
Oxford  OX1 3NP, United Kingdom}
\date{\today}
\maketitle

\begin{abstract}
We consider non-interacting bosonic excitations in disordered systems,
emphasising generic features of quadratic Hamiltonians in the absence of Goldstone modes.
We discuss relationships between such Hamiltonians and the symmetry classes
established
for fermionic systems. We examine the density $\rho(\omega)$
of excitation frequencies $\omega$, showing how the universal behavior
$\rho(\omega) \sim \omega^4$ for small $\omega$ can be obtained both from general arguments and
by detailed calculations for one-dimensional models.
%
%
\end{abstract}

]
\vskip2pc

An understanding of universal properties of excitations in disordered systems
occupies a central place in condensed matter physics. Much of the work in this
area has focussed on systems of non-interacting fermions, as models
for quasiparticles in disordered conductors or superconductors
with interactions treated at the mean-field level. Classification
of such systems according to symmetry provides an important starting point,
and in many instances the possibilities are the three represented by the  Wigner-Dyson random matrix ensembles \cite{Mehta}.
A recent development, however, has been the recognition that there exist symmetry classes additional to those of
the Wigner-Dyson ensembles. These additional symmetry classes arise in fermionic systems which have either chiral symmetry,
as for tight-binding models on bipartite lattices with only off-diagonal
disorder \cite{Gade}, or particle-hole symmetry, as for the Bogoliubov - de Gennes
Hamiltonian in disordered superconductors \cite{Altland-Zirnbauer}.
A characteristic feature of both cases is that eigenstates at positive and negative energies are related in
pairs and zero energy emerges as a special point in the spectrum.

It is natural to anticipate that similar mathematical structure may be important
for the theory of non-interacting bosonic excitations or classical harmonic
modes in disordered systems. Our aim in the following is to examine
how far this is the case and what consequences it has.
Studies of systems of this type have a long history,
with celebrated early work by Dyson \cite{Dyson} on the dynamics of a
disordered chain of masses and springs, and applications which include
phonons in disordered solids \cite{GKK} and spin waves in random magnets \cite{magnon-review}.
An obvious parallel between between excitations in these systems and those in fermionic systems belonging to the one of the additional symmetry classes
is that bosonic excitations arise in positive and negative frequency pairs,
and zero is a special point in the frequency spectrum. A second,
more formal parallel is that random magnets, for example,
in common with superconductors, may give rise to quadratic Hamiltonians containing
terms that annihilate and create particle pairs. There are also
clear differences between systems with bosonic and fermionic excitations,
which prevent an elementary transcription of established ideas.
Most importantly, while the exclusion principle guarantees that a fermionic
system with a quadratic Hamiltonian has a ground state, for a corresponding bosonic
system constraints must be imposed to ensure that it is stable.
As a consequence, even to construct a phenomenological treatment
such as random matrix theory for bosonic excitations, it is necessary
to keep in mind their origins in an appropriate non-linear problem.
In turn, a distinction arises that is specific to bosonic excitations,
between those which are Goldstone modes and those which are not.
We focus below on the latter and discuss elsewhere systems
with continuous symmetry \cite{long}.

Any quadratic bosonic Hamiltonian can be written in the forms
\begin{eqnarray}
\label{hamreal}
  H&=& \sum_{i,j=1}^N \left[ M_{ij} p_i p_j +  K_{ij} x_i x_j +  C_{ij} (x_i p_j + p_j x_i) \right] \nonumber \\
&\equiv&\sum_{i,j=1}^{2N} \cH_{ij} \xi_i \xi_j \,\equiv\,
\frac{1}{2}({\bf a}^{\dagger}\, {\bf a})\left( \begin{matrix} \Gamma & \Delta\\
\Delta^{\dagger}& \Gamma^T \end{matrix} \right) \left(\begin{matrix} {\bf a}\\ {\bf a}^{\dagger}\end{matrix}\right)\,.
\end{eqnarray}
Here, $x_i$ and $p_i$ are the coordinates
and momenta of the oscillators, $\xi_i=p_i$, $\xi_{N+i} = x_i$ with $1 \le i \le N$, and
the vectors ${\bf a}^{\dagger}$, ${\bf a}$ have 
as entries bosonic creation and annihilation operators $a^{\dagger}_i, a_i=(x_i\pm ip_i)/\sqrt{2}$.
The matrix $\cH$ is real symmetric, $\Gamma$ is Hermitian and $\Delta$
is symmetric. The condition for time-reversal invariance is that $C=0$
or, equivalently, that $\Gamma$ and $\Delta$ are real.

We are concerned with spectral properties of
models of this kind which are generic when $\cal H$ is random. One anticipates
that these will be found at small excitation frequencies $\omega_i$,
and we concentrate particularly on their average density
$\rho(\omega)=N^{-1}\langle\sum\delta(\omega - \omega_i)\rangle$ at small $\omega$.
Indeed, the final expression of Eq.\,\rf{hamreal} illustrates the parallel between bosonic models and the Bogoliubov - de Gennes Hamiltonian,
from which universal behaviour has been derived for the density of
quasiparticle states
in disordered, gapless superconductors
\cite{Altland-Zirnbauer}. In fermionic systems such behaviour arises essentially from the interplay of disorder and symmetry,
and does not depend on details of the distribution
of Hamiltonian matrix elements. By contrast, for the bosonic systems we discuss, we show that the requirements of stability impose features on the
distribution of $\cal H$
which determine the form of $\rho(\omega)$ at small $\omega$.

A simple argument leading to this conclusion
has been given previously \cite{IKP,R,Fogler}. In
outline it is as follows.
Recall that the Hamiltonian of  Eq.\,(\ref{hamreal}) is
characterised not only by the oscillator
frequencies $\omega_i$ but also by the
eigenvalues $\kappa_n$ of the matrix $\cal H$, which we refer to as oscillator
{\it stiffnesses}; let $d(\kappa)=(2N)^{-1}
\langle\sum \delta(\kappa-\kappa_n)\rangle$ be their average density.
From a discussion of the
curvature distribution
at absolute minima of random functions of one variable, it is
suggested \cite{IKP} that $d(\kappa)\propto \kappa^{3/2}$ for small $\kappa$.
Then, using the relation $\omega_i^2 = \kappa_1 \kappa_2$ which holds in
a single-mode system, one arrives at
$\rho(\omega) \propto d(\omega^2)\, \omega \propto \omega^4$
for small $\omega$.
A weakness at both steps in this argument is that one
degree of freedom is treated in isolation. We show here how
the same behaviour emerges without such a restriction.

In a stable system, stiffnesses are positive and frequencies are
real. It is helpful to introduce a description that guarantees this
property. To this end, write $\cH$ as a square of a real matrix $Q$,
in the form $\cH=Q^T Q$ (possible provided all $E_n \geq 0$).
Also, note that frequencies are
the eigenvalues of an auxilliary matrix $\cH'=\sigma_2 \cH$,
where $\sigma_2=\sigma_y \otimes I_N$,  $\sigma_y$ is the usual Pauli matrix and
$I_N$ is the $N \times N$ identity matrix \cite{Blaizot}. Since the eigenvalues of
$\sigma_2 Q^T Q$ coincide
with those of $\Omega=Q \sigma_2 Q^T$,
which is Hermitian and antisymmetric,
frequencies are real and occur in pairs $\pm \omega_i$.
While there is in general no simple relation between
stiffnesses and frequencies, several special cases represent important
exceptions. If (in Eq.\,(\ref{hamreal})) $M=I_N$ and $C=0$
(as for vibrational problems with all masses equal), then $\omega_i = \pm \sqrt{\kappa_i}$,
where the non-trivial stiffnesses appearing here are the eigenvalues of $K$;
whereas if $M=K$ and $C=0$ (as for a random-bond ferromagnet), then $\omega_i= \pm \kappa_i$.

Moving beyond these special cases,
a straightforward approach to multimode problems comes from concentrating on
disorder realisations in which one stiffness, say $\kappa_1$, is much smaller than all
others. The probability density for such disorder realisations
varies as $\kappa_1^{3/2}$, by the arguments of Ref\,\cite{IKP}.
In the limiting case $\kappa_1=0$, two eigenvalues of
$\Omega$ vanish (since ${\rm det}(\cH) = {\rm det} (\Omega)$,
and since the eigenvalues of $\Omega$ are paired), while for
$0 < \kappa_1 \ll \kappa_n$, $n \not= 1$, perturbation theory
yields a pair of excitation frequencies $\omega_i \propto \pm \sqrt{\kappa_1}$.
As a result, $\rho(\omega) \propto \omega^4$ for $\omega \ll \omega_p$, where
$\omega_p$ is the lowest excitation
frequency in a typical sample. More generally, consider a macroscopic system in which
low frequency modes are localized with a localisation length
which remains finite as $\omega \to 0$.
Treating each localization volume independently,
we obtain $\rho(\omega) \propto \omega^4$ for $\omega \ll \omega_p$, where
$\omega_p$ is in this case the lowest
excitation frequency in a typical localization volume.

Turning to detailed calculations, it is useful
to construct a formulation in which $Q$ appears linearly, by considering
the enlarged matrices
\begin{equation}
\label{chi}
\t \cH = \left( \begin{matrix} 0 & Q \cr Q^T & 0 \end{matrix} \right)
\hspace{0.2cm}\hbox{\rm and}\hspace{0.2cm}
\t \cH' =  \left( \begin{matrix} 0 & Q \cr \sigma_2 Q^T & 0 \end{matrix} \right) \,.
\end{equation}
Clearly, the eigenvalues of $\cH$ and  $\cH'$ are squares of those of $\t \cH$  and $\t \cH'$ respectively.
The $2\times 2$ structure of $\t \cH$ displayed in Eq\,(\ref{chi})
is interesting partly because it is the defining feature of the
chiral symmetry classes, studied previously in a variety of contexts \cite{Gade,Eggarter,Comtet,Verbaarshot,Slevin,Brouwer},
while matrices with the structure of $\t \cH'$ constitute a new, chiral bosonic problem.
Whereas much past work on chiral symmetry classes has taken the elements of $Q$ to be
independent random variables, drawn from a given distribution, it is clear following our introductory discussion
that a central concern in our case is to determine the distribution
of $\cH$ and hence that of $Q$. Moreover, to allow further progress, the form of $Q$ must be
sufficiently simple (for example, local), and it does not seem obvious in advance
whether this will be the case for any given problem. In fact, there
is considerable flexibility, since $\cH$ determines
$Q$ only up to left multiplication by an orthogonal matrix,
and we have found convenient, explicit expressions for $Q$
in several one-dimensional examples \cite{long}. In addition, the approach proves useful
even for problems in which $Q$ is known only implicitly, as we now show.

We consider the random field XY model, which has been studied extensively as
a description of pinned charge density waves \cite{FL}.
It has the Hamiltonian
\begin{equation}
\label{ener}
H= \int_0^L \,\left[ {1 \over 2} 
\Pi^2 + {1 \over 2} (\partial_x \phi)^2 + h(\phi,x) \right]\,dx
\end{equation}
where $\Pi$ is the momentum conjugate to the angle
$\phi$, $h(\phi,x) = h(x) \cos (\phi(x)-\chi(x))$,
and $h(x)$ and $\chi(x)$ are the random
amplitude and phase of an applied field, with $\VEV{h(\phi,x)}= 0$ and
$\VEV{h(\phi,x) h(\phi', x')}= \delta(x-x')h_0 \cos
(\phi-\phi') $.
This continuum problem has, in the discrete notation used above,
$M=I_N$, $C=0$ and hence $\omega=\pm \sqrt{\kappa_i}$.
We are therefore concerned only with $K$:
for the sake of an obvious analogy, we refer to it as
a Hamiltonian, denoting it by $\cH$ and its eigenvalues by $E$ in place of $\kappa$.

Stationary configurations of the field $\phi$ satisfy
\begin{equation}
\label{eqm}
-\partial^2_x\phi +\partial_\phi h(\phi,x) = 0
\end{equation}
while amplitudes $\psi$ of normal mode excitations about the ground-state $\phi_0$,
with frequency $\omega = \pm \sqrt{E}$, obey
\begin{equation}
\label{CDW}
\cH \psi  \equiv - \partial_x^2\psi + \partial^2_\phi h(\phi_0,x) \psi = E \psi  \,.
\end{equation}
We want to find the density of states for the Hamiltonian of Eq.\,\rf{CDW}.
In this
form, this problem
has been the subject of many publications \cite{FL,Fei,CDW}.
Summarising what is known: a characteristic energy scale, the pinning
energy \cite{Larkin} $E_p$, separates two regimes. For $E\gg E_p$
the density of states is only weakly affected by the random field and can be computed
perturbatively in the field strength. For $E\ll E_p$
the density of states is strongly influenced by disorder and
believed to vary with $E$ as a power law, $d(E) \propto E^\beta$, but the
value of $\beta$ remains controversial\cite{FL,R,Fogler}.
Here we show, in argeement with arguments due to Aleiner and Ruzin \cite{R} and to Fogler \cite{Fogler},
that $\beta=3/2$.

Since Eq.\,\rf{CDW} has the form of a Schr\"odinger equation with a random potential
$\partial^2_\phi h(\phi_0,x)$, it is tempting to anticipate behaviour familiar
from simple choices of potential distribution, such as Gaussian white noise.
That would be too naive, however, because --
as for any bosonic Hamiltonian -- the spectrum of $\cH$ is positive and so the potential
cannot be arbitrarily random. To investigate the implications of this fact, it is
useful to follow Feigelman\cite{Fei}, introducing
the notion of a partial energy, $\cE(\phi,y)$: the ground-state energy of the half-chain
with coordinate values $0\leq x\leq y$ and boundary condition $\phi(y)=\phi$.
Interpeting Eq.\,\rf{eqm} as an equation of motion for a system with space
coordinate $\phi$ and time coordinate $x$, $\cE(\phi,y)$ plays the role of an action
and satisfies the Hamilton-Jacobi equation
\begin{equation}
\label{KPZ}
\partial_x \cE + \oh \left( \partial_\phi \cE \right)^2 = h(\phi,x)
\end{equation}
while the classical trajectory $\phi_0(x)$ obeys $d \phi_0/dx = \partial_\phi \cE$.
Now define $V(x)=\partial^2_\phi \cE(\phi_0(x),x)$, which we call the chiral potential.
As a direct consequence of Eq.\,\rf{KPZ} it
satisfies
\begin{equation}
\label{slope}
\dd{x}V + V^2 = \partial^2_\phi h(\phi_0(x),x)\,.
\end{equation}
The Hamiltonian of Eq.\,\rf{CDW} can therefore be rewritten as $\cH=Q^T Q$, where
\begin{equation}
\label{1dchiral}
Q=-\dd{x}+V(x)\,.
\end{equation}
Thus the bosonic problem of Eq.\,\rf{CDW} is
equivalent to a one-dimensional chiral problem specified by Eqs.\,\rf{chi} and \rf{1dchiral},
in which $V(x)$ should be determined along with $\phi_0(x)$.

At this point we can draw on the work of Comtet {\sl et al}\cite{Comtet}, in which
one-dimensional chiral problems of this kind have been analyzed in detail. In particular,
they show that
if the chiral potential $V(x)$ has a positive average $\VEV{V(x)} = E_p^{1/2}$ (sometimes
referred to as staggering \cite{Brouwer}), then
the low-lying ($E \ll E_p$)
states of Eq.\,\rf{chi} are localized with a localization length $\xi \sim E_p^{-1/2}$.
In this regime,
the integrated density of states
$N(E)=\int_0^E\,d(E') dE'$
can be calculated as the probability of a negative fluctuation of $V(x)$ in the interval $x_1<x<x_2$
for which $2\int_{x_1}^{x_2} dx~V(x) \equiv 2U < \log(E)$.
The probability for a rare event of this kind is expected under many circumstances to
vary with $E$ as $\exp[\alpha \log(E)]=
E^\alpha$, but the value of $\alpha$ depends on the details of the distribution
of $V(x)$.
In the following, we establish the equivalence implied by our notation, between  $\VEV{V(x)}^2$ and pinning
energy, and calculate the low-lying density of states for the random field XY model
by studying rare negative fluctuations of the particular chiral potential
that arises in this context.

To appreciate why $\VEV{V(x)}>0$, consider the ground state of the half-chain as a function
of the boundary condition, $\phi$: it varies smoothly except for jumps at a small number of
isolated points within one period. At these points $\cE(\phi,y)$ is continuous but has an
unward cusp \cite{Fogler}; since $\cE(\phi,y)$ is periodic, an average of $V(x) \equiv \partial_\phi^2 \cE$
computed excluding these cusps (as on $\phi_0$) is naturally positive, and its size
can be estimated from the average sum of discontinuities in $\partial_\phi \cE$. Alternatively,
one can attempt a more detailed analysis of Eq.\,\rf{KPZ}, which is similar to the
KPZ equation \cite{KPZ}, but with two differences. One, the absence of a diffusion term $D \partial_x^2\cE$,
is unimportant since it is well known that the limit $D \to 0$ in the KPZ equation
yields solutions which are the global minimum of Eq.\,\rf{ener}.
The second, involving the nature of noise correlations is more significant:
while studies of the KPZ equation deal with noise that is uncorrelated in
both $x$ and $\phi$, we are concerned with noise that has long-range correlations
in $\phi$. This situation is familiar in the context of Burgers turbulence \cite{BT},
and indeed, introducing $u=\partial_\phi \cE$ one finds the Burgers equation
$
u_x + u u_\phi = \partial_\phi h.
$
Cusps in $\cE(\phi,y)$ as a function of $\phi$ corrspond to the shock waves of
Burgers turbulence, and using dimensional analysis the dependence on noise strength $\VEV{V(x)}\sim 
h_0^{1/3}$ can be
obtained from Eq.\,\rf{KPZ}, as in the standard picture of the pinning energy.

Turning to fluctuations in $V(x)$, a direct attack, via a solution of Eq.\,\rf{KPZ}, is not appropriate
since many aspects of Burgers turbulence remain controversial. Instead we treat Eq.\,\rf{slope} as a Langevin
equation for $V(x)$ with $\partial^2 h(\phi_0(x),x)$ playing the role of a random force.
If $\phi_0(x)$ is determined by minimising Eq.\,\rf{ener} for fixed $\phi(0)$ and $\phi(L)$,
the resulting random force has built-in correlations which ensure that $V(x)$
remains bounded. Alternatively, one may solve Eq.\,\rf{eqm} for $\phi(x)$ with
fixed initial conditions, $\phi(0)$ and $\partial_x \phi(0)$. In this case, the
random force is uncorrelated and appropriate values for $\partial_x \phi(0)$ generate
ground states corresponding to different choices of $\phi(L)$. Other values of
$\partial_x \phi(0)$, however, correspond to maxima of Eq.\,\rf{ener}, and still others
to higher energy local minima.
From the perspective of
the Langevin equation, a maximum of Eq.\,\rf{ener} is signalled by escape of $V(x)$ towards large negative values.
Those noise realisations that result in escape should be eliminated by supplementing Eq.\,\rf{slope}
with an absorbing boundary condition at
$V(x)=-\infty$. The trajectories of $V(x)$ that are not absorbed
should be weighted in two ways. First, we must eliminate solutions to  Eq.\,\rf{eqm}
which are only local minima of Eq.\,\rf{ener}. They are separated
from the absolute minimum by maxima and hence have
neighboring trajectories on which $V(x)$ escapes. The necessary weight is therefore
the probability that no near neighbors of a given trajectory are absorbed.
Second, we weight trajectories representing the absolute minimum for
different $\partial_x \phi(0)$ in such a way as to give a uniform density on
$\phi(L)$.

Omitting in the first instance both these weights, the form of the probability distribution
$P(U)$ at large negative $U$ can be found by passing from the
Langevin equation to a path integral, fixing $U$ with a Lagrange multiplier,
and calculating the lowest eigenvalue of the corresponding Fokker-Planck operator,
which is non-zero because of the absorbing boundary condition.
In this way we obtain $P(U)\sim e^{2U}$ as $U \to -\infty$.
To calculate the weighting factors,
we consider a family of trajectories parameterized by $s$, with
coordinate $\phi(x) + \eta(s,x)$ and chiral potential
$V(x) + W(s,x)$, where $\eta(s,x)$ and $W(s,x)$ are assumed small.
Deriving linearized evolution equations by expanding Eqs.\,\rf{eqm} and \rf{slope},
and solving these, we find $\eta(s,x_2)=e^U \eta(s,x_1)$ and $W(s,x_2)=e^{-2U}F(s)$,
where
\begin{eqnarray}
\label{F}
   F(s)=&&W(s,x_1) \nonumber\\
   + \eta(s,x_1)&&\int_{x_1}^{x_2}\exp\left(3\int_{x_1}^{x}V(y)dy\right)\partial^3_\phi h(\phi,x) dx\,.
\end{eqnarray}
In order that no trajectories neighboring $\phi(x)$ are absorbed, we
require $W(s,x_2)\gtrsim -1$ for all $s$ and
hence, at large negative $U$, $F(s)\gtrsim -e^{2U}$, which occurs with probability $e^U$.
Of the surviving trajectories, we should retain only those that have a fluctuation in the
integrated chiral potential similar to $U$. These satisfy the condition $W(s,x_2)\lesssim 1$,
and hence $F(s)\lesssim e^{2U}$, implying $|\eta(s,x_1)|\lesssim e^U$ and  $|\eta(s,x_2)|\lesssim e^{2U}$,
thus generating a weight $e^{2U}$. Combining these factors, the probability density that $U$ is large
and negative on $\phi_0(x)$ varies as $e^{5U}$, yielding
$N(E) \sim E^{5/2}$, and hence $\rho(\omega)\sim \omega^4$,
as anticipated above. A detailed account of these arguments
will be presented elesewhere \cite{long}.

The mapping described here of Eq.\,\rf{ener} into a chiral problem also goes through
for a discrete version of the XY spin chain. It gives,
not surprisingly, a staggered one-dimensional random hopping model, as studied for example in
Refs.\,\cite{Eggarter,Brouwer}.
This formulation is especially well suited to numerical calculations using the transfer matrix technique.
Fig.\,1 shows the result of
a computation of $N(E)$ for a
XY spin chain of $10^6$ sites, with Hamiltonian
$H=\sum_i \left\{ h_i(\phi_i)-\cos(\phi_i-\phi_{i-1})
\right\}$, where  $h_i(\phi)=A_i \cos(\phi-\chi_i )$,
and $[A \cos(\chi), A \sin(\chi)]$ is uniformly
distributed on a disc of radius $0.01$.
The power law obtained analyticially is confirmed.

\begin{figure}[htbp]
\centerline{\epsfxsize=2.0in \epsfxsize=3.0in
\epsfbox{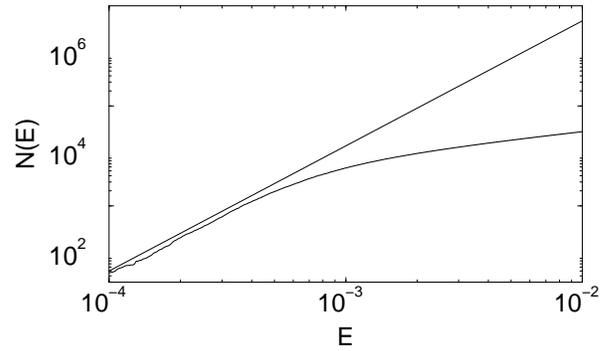} }
\caption{The integrated density of stiffnesses, compared with the theoretically
predicted power, $\beta + 1 = 5/2$}
\end{figure}

We remark finally that, while the random field XY chain represents a special case in the sense
noted above, since frequencies and stiffnesses are simply related, we expect the same form more generally for
$\rho(\omega)$, as implied by the discussion preceeding Eq.\,\rf{chi}. Calculations
for the random field Heisenberg spin chain lead to two-channel chiral problems involving $\t \cH$ and $\t \cH'$.
For such problems we obtain \cite{long}
$d(\kappa)\sim \kappa^{3/2}$ and $\rho(\omega)\sim \omega^4$.

We thank A. Altland and M. Zirnbauer for valuable discussions,
and acknowledge support by EPSRC under grant GR/J78327.

\end{document}